\documentclass[twocolumn]{article}
\usepackage[utf8]{inputenc}
\usepackage{amsmath}
\usepackage{amsfonts}
\usepackage{amssymb}
\usepackage{graphicx}
\usepackage{booktabs}
\usepackage{array}
\usepackage{multirow}
\usepackage[table]{xcolor}
\usepackage{listingsutf8}
\usepackage{tikz}
\usepackage{adjustbox}
\usepackage{geometry}
\usepackage{hyperref}
\usepackage{float}
\usepackage{microtype}
\usepackage{tcolorbox}
\definecolor{codegray}{rgb}{0.5,0.5,0.5}
\definecolor{codepurple}{rgb}{0.58,0,0.82}
\definecolor{backcolour}{rgb}{0.95,0.95,0.92}
\definecolor{keywords}{rgb}{0.26, 0.44, 0.56}
\definecolor{strings}{rgb}{0.31, 0.56, 0.16}
\definecolor{comments}{rgb}{0.35, 0.45, 0.35}

\lstdefinestyle{pythonstyle}{
  backgroundcolor=\color{backcolour},
  commentstyle=\color{comments}\ttfamily\itshape,
  keywordstyle=\color{keywords}\bfseries,
  numberstyle=\tiny\color{codegray},
  stringstyle=\color{strings},
  basicstyle=\ttfamily\footnotesize,
  breakatwhitespace=false,         
  breaklines=true,                 
  captionpos=b,                    
  keepspaces=true,                 
  numbers=left,                    
  numbersep=5pt,                  
  showspaces=false,                
  showstringspaces=false,
  showtabs=false,                  
  tabsize=4,
  language=Python
}
\title{Factor Engine: A Python Library for Systematic Financial Factor Computation and Analysis}
\author{Ata Keskin \\ Technical University of Munich}
\date{July 27, 2025}

\begin{document}

\maketitle

\begin{abstract}
    Factor Engine is a high-performance, open-source Python library designed for the systematic computation and analysis of financial factors. Built around a modular and extensible API that leverages Python decorators, Factor Engine enables users to define custom factors with ease and integrates seamlessly with the modern data science ecosystem. To assess its practical effectiveness, we compare the mispricing factors computed by Factor Engine to those generated using a reference Stata implementation, finding that both approaches yield highly similar results and comparable performance in backtesting analyses. Furthermore, we experimentally apply these factors within machine learning workflows for trading strategy development, illustrating their practical utility and potential for quantitative finance research.
\end{abstract}

\section{Introduction}
\subsection{Motivation}
Factor investing, rooted in the Arbitrage Pricing Theory (APT) \cite{burmeister1986} and the Fama-French multi-factor models \cite{fama1992}, is a cornerstone of modern quantitative finance. These models posit that asset returns can be explained by their exposure to systematic risk factors, such as size, value, momentum, and quality \cite{carhart1997}. The ability to reliably compute these factors from raw financial data is crucial for academic research, risk management, and the development of "smart beta" investment strategies.

However, practitioners often face challenges with existing tools. Traditional software like SAS or Stata can be costly and may not integrate well with the modern Python data science stack. While the Python ecosystem offers powerful libraries for data analysis (pandas \cite{mckinney2010pandas}, polars \cite{polars2023}, NumPy \cite{harris2020array}) and machine learning (scikit-learn \cite{pedregosa2011scikit}, PyTorch \cite{paszke2019pytorch}), a dedicated, high-performance tool for financial factor engineering is less established. Existing financial libraries often focus on data acquisition or backtesting, leaving a gap for a specialized, transparent, and efficient factor computation engine.

\subsection{Design Goals}
Factor Engine aims to fill this gap by providing a library that is, by design, performant and scalable, capable of handling large-scale panel datasets covering thousands of assets over decades without performance bottlenecks. It is also built to be transparent and reproducible, offering clear, well-documented implementations of established financial factors, which is critical for academic and practical validation. A core goal is extensibility; the library provides a modular and intuitive architecture that allows researchers to easily define and test their own custom factors with minimal boilerplate code, fostering innovation. Finally, the tool is designed to be seamlessly integrated into the existing Python ecosystem, working smoothly with common libraries for financial analysis and machine learning.

\section{Background and Related Work}
Financial factors are characteristics of securities that help explain their risk and return. Seminal works by Fama and French (1992) identified size (SMB) and value (HML) as key factors, while Carhart (1997) added momentum (MOM). Since then, more key factors have emerged, including profitability, investment, and low volatility.

Several open-source Python libraries address parts of the quantitative workflow. \texttt{pyfolio} \cite{pyfolio} and \texttt{Alphalens} \cite{alphalens} provide excellent tools for performance and factor analysis but assume the factors are already computed. Libraries like \texttt{Zipline} \cite{zipline} and \texttt{bt} \cite{bt} are powerful backtesting engines. Factor Engine is complementary to these tools, focusing specifically on the efficient and reliable computation of the factors themselves, which can then be fed into these other libraries for analysis and backtesting. Unlike general-purpose tools, Factor Engine is purpose-built for the nuances of financial factor calculation, such as handling time-series operations (lags) and ensuring point-in-time correctness.

\section{System Design and Architecture}

\subsection{Design Philosophy}
The design philosophy of Factor Engine is centered on three core principles. First is modularity, where factors are defined using standalone functions, decoupled from the core engine. This makes the system easy to maintain, test, and extend without affecting other components. Second is compatibility; while the library is built on Polars for its superior performance, it seamlessly integrates with other DataFrame libraries, allowing users to leverage the vast and familiar Python data science ecosystem without friction. The third and most crucial principle is extensibility. The library is engineered to empower users to add custom factors with minimal effort, a goal achieved through a powerful and intuitive decorator-based API that abstracts away the underlying complexity.

\subsection{Internal Workflow}
The engine follows a systematic process to transform raw panel data into computed factors, ensuring correctness and high performance.

\subsubsection{Input Data Processing}
The workflow begins when a user initializes a Factor Engine instance with a DataFrame. Together with the DataFrame provided, the constructor expects a list of column name identifiers. The indentifiers \textit{id\_col} (for the asset identifier) and \textit{date\_col} (for the time-series index) are necessary. It then maps the provided column names (e.g., \textit{total\_assets\_col}="WC02999") to a standardized internal naming convention (e.g., `ta`) for consistency. This mapping provides flexibility, allowing the engine to work with any data schema as long as the required columns are specified. The date column is immediately cast to a datetime type to enable time-series operations.

\subsubsection{Lagged Column Management}
At the heart of the engine's ability to handle time-series data correctly is the class \textbf{OffsetColumnManager}. When a factor's formula requests a lagged column, for example, the market value from 12 months prior using \texttt{self.col("mv", "12mo")}, the operation is not executed immediately. Instead, the \textbf{OffsetColumnManager} instance registers a request for the `mv` column with a 12-month lag.

The actual creation of these lagged columns is deferred until the \texttt{compute} method is called. Just before a factor is calculated, the manager's \texttt{compute\_offset\_data} method is invoked. It efficiently generates all unique lagged columns required for the computation in a single pass, using the utility method \texttt{join\_with\_offset} which performs an as-of join on the underlying DataFrame with itself. Lagged columns with the same lag value are computed in the same call to \texttt{join\_with\_offset}. This deferred, bulk-processing approach is significantly more performant than calculating lags individually.

\subsubsection{The Decorator-Based API}

The library is designed so that users can add, modify, or remove factors with minimal effort and without touching the core engine logic. This is achieved by leveraging Python's support for first-class functions and decorators. Concretely, we implement the following decorators: \texttt{@simple\_factor} and \texttt{@advanced\_factor}, which automatically register any function they adorn as a computable factor in the engine's registry.

\paragraph{\texttt{@simple\_factor}} is designed for factors that can be defined as a single Polars expression. The user writes a function that returns this expression. The decorator's wrapper logic then handles the complex boilerplate: it checks for data requirements, invokes the \textbf{OffsetColumnManager} to prepare lagged data, applies the user's expression to the DataFrame, and handles optional post-processing like z-scoring. The decorated function itself gets turned into a function returning a Polars DataFrame instead of an expression.

\paragraph{\texttt{@advanced\_factor}} serves as a powerful alternative for complex factors that cannot be expressed in a single line. A function decorated with \texttt{@advanced\_factor} receives the full historical DataFrame. This gives the developer complete control to implement stateful logic, rolling window calculations, or operations involving external datasets. The decorated function is expected to return a DataFrame consisting of 3 columns: the asset id, the date, and the computed factor. \\

In Python, functions are first-class objects, meaning they can be passed around, stored in data structures, and dynamically modified. Factor Engine exploits this by using decorators to register factor functions and wrap them with additional logic. When a user writes a new factor as a function and decorates it with \texttt{@simple\_factor} or \texttt{@advanced\_factor}, the decorator intercepts the function definition, attaches metadata, and adds it to the engine's internal registry. This registry is then used by the engine to discover and execute all available factors.

The decorator mechanism also allows the engine to inject boilerplate code (such as data validation, lagged column management, and post-processing) without requiring the user to write it themselves. This separation of concerns means that users only need to focus on the mathematical definition of their factor, while the engine handles all the technical details of computation and integration.

Because decorators are just functions that take other functions as arguments, they can be composed, extended, or replaced as needed. This makes the API highly flexible and future-proof. For example, new decorators could be introduced to support GPU acceleration, distributed computation, or custom logging, all without changing the user's factor definitions.

\subsubsection{The \texttt{compute} Method}
The \texttt{compute} method is the primary user entry point for executing bulk factor calculations. A user can request one or more factors by name. The method iterates through the requested factors, looks them up in the factor registry, and executes the corresponding decorated function. As each factor is computed, its results (a DataFrame containing the id, date, and factor value) are joined back to the main data table. This process incrementally builds the final output DataFrame containing all the requested factors, which is then returned to the user. The method also gracefully handles cases where a factor's data requirements are not met, skipping it with a warning instead of raising an error. 

\subsection{Data Quality and Preprocessing}
The engine provides built-in, configurable mechanisms to solve common data quality problems in financial datasets. It handles missing data through forward- or backward-filling strategies applied independently within each asset's history, preventing data leakage across securities. Users can supply input data at any reporting frequency. Factor Engine resamples the data internally to match the frequency specified for metric computations. This design allows users to work with columns lagged by any amount, without needing to manage alignment or frequency themselves. The \textbf{OffsetColumnManager} aligns all lagged data to the correct timestamps and enforces strict prevention of look-ahead bias, which is a critical error where future information influences past values.

The library also includes utility methods for preprocessing. It offers functions to read data directly from Excel sheets or Stata files. There are also some statistical preprocessing methods present. These include a resampling method and a winsorization method intended to be used when manually constructing a factor via the decorator \texttt{@advanced\_factor}. Winsorization helps reduce the influence of extreme values that can distort statistical analysis and model training. The \texttt{@simple\_factor} decorator features a built-in option to automatically winsorize data. Users can specify upper and lower quantiles for this winsorization step, which takes place before factor computation.

\section{Use Case: \textbf{MispricingFactors}}

The \texttt{mispricing.py} module contains a reference implementation that showcases how to compute the standard set of 11 mispricing factors introduced by Stambaugh and Yuan (\cite{stambaugh_yuan_2017}) using the Factor Engine framework. At its core, the module defines the \textbf{MispricingFactors} class, which extends Factor Engine and leverages its powerful decorator-based API to simplify the creation and management of factor computations. Each of the 11 mispricing factors are implemented as a dedicated methods within the class, with decorators responsible for automatic registration of the factors. This design not only enhances code readability but also makes it easy to extend or customize factors based on specific research needs. As a result, \texttt{mispricing.py} serves both as a practical tool for factor calculation and a clear example of how to effectively utilize Factor Engine’s architecture for building complex financial metrics. Each factor is implemented as a method, leveraging the engine's decorator-based API for clarity and extensibility.

Below is a brief description of each mispricing factor:

\begin{itemize}
  \item[] \textbf{Net Stock Issues} \\
  Measures the annual log change in split-adjusted shares outstanding. Based on the idea that firms issue equity when overvalued, this factor captures overpricing due to sentiment-driven issuance behavior.

  \item[] \textbf{Composite Equity Issues} \\
  Captures the change in market capitalization not attributable to returns. Calculated as the 12-month change in market equity minus the 12-month stock return, this measure identifies firms whose valuation increase is not fully explained by performance.

  \item[] \textbf{Accruals} \\
  Measures the ratio of accruals (change in noncash working capital minus depreciation) to average total assets. High accrual firms tend to underperform due to investor misperceptions about the persistence of accrual-based earnings.

  \item[] \textbf{Net Operating Assets} \\
  Represents the balance sheet difference between operating assets and liabilities, scaled by lagged total assets. High net operating assets suggest investor neglect of cash flow information, predicting lower returns.

  \item[] \textbf{Asset Growth} \\
  Defined as the year-over-year growth rate in total assets. Firms with high asset growth tend to be overvalued, as investors overreact to expansion signals, leading to lower future returns.

  \item[] \textbf{Investment-to-Assets} \\
  Captures overinvestment by comparing annual changes in gross property, plant, equipment, and inventory to lagged total assets. High investment predicts underperformance due to managerial overconfidence and investor underreaction.

  \item[] \textbf{Distress (Failure Probability)} \\
  Estimated via a dynamic logit model incorporating returns, volatility, size, leverage, and other financial indicators. High distress probability is associated with lower future returns, contradicting rational risk-based pricing.

  \item[] \textbf{O-score} \\
  A static distress indicator based on Ohlson (\cite{ohlson1980}), combining several accounting ratios to estimate bankruptcy probability. Higher O-scores imply greater financial risk and are associated with lower expected returns.

  \item[] \textbf{Momentum} \\
  Based on past returns from month $t{-}12$ to $t{-}2$. Firms with strong recent performance tend to continue outperforming, while poor performers tend to lag—an effect well-established in asset pricing literature.

  \item[] \textbf{Gross Profitability} \\
  Measured as gross profit over assets, this factor emphasizes operating efficiency and core profitability. More profitable firms tend to yield higher risk-adjusted returns.

  \item[] \textbf{Return on Assets} \\
  Calculated as quarterly income before extraordinary items over lagged total assets. Higher return on assets signals greater operational performance and predicts stronger future returns.
\end{itemize}

The class \textbf{MispricingFactors} showcases how and when the decorators introduced above should be used to implement different factors. For instance, a straightforward metric like Return on Assets (ROA) is implemented as a \texttt{@simple\_factor}. The implementation is concise, as the user only needs to define the core logic:
\begin{tcolorbox}[title=Simple Factor Example]
\begin{lstlisting}[style=pythonstyle]
@simple_factor(name="roa",
               requires=["ibq", "ta"])
def roa(self, lag=3) -> pl.Expr:
    return (self.col("ibq") 
         / self.col("ta", f"{lag}mo"))
\end{lstlisting}
\end{tcolorbox}

In contrast, the Distress factor is implemented as an \texttt{@advanced\_factor}. This is necessary for its complex, multi-step computation which involves rolling averages and external market data. The decorator provides the full DataFrame history to the function, giving the developer complete control for stateful calculations. 

This decorator also allows for additional keyword arguments to be passed to the method such as a DataFrame containing historic S\&P500 data in this case. This clear separation of concerns allows the library to be both highly efficient for simple cases and powerful enough for complex, multi-step factor definitions.

After the factors have been implemented, the user can simply instantiate \textbf{MispricingFactors} with data from any source and compute the registered factors.
\begin{tcolorbox}[title=Example Workflow]
\begin{lstlisting}[style=pythonstyle]
import polars as pl
from factor_engine import MispricingFactors
from factor_engine.utils import read_excel

# 1. Load data
df = read_excel("financial_data.xlsx")
sp500 = pl.read_parquet("sp500.parquet")

# 2. Initialize engine with data and schema
engine = MispricingFactors(
    data=df,
    id_col="ticker",
    date_col="date",
    # ... other column mappings
)

# 3. Compute all mispricing factors (include sp500 for the Distress factor)
all_factors = engine.compute(sp500=sp500)

# 4. Compute a single factor
momentum = engine.compute("momentum", lag=6)
# Or
momentum = engine.momentum(lag=6)
\end{lstlisting}
\end{tcolorbox}
The resulting DataFrame of factors can then be used in a backtesting library like \texttt{Zipline} or for training a machine learning model, as demonstrated in the next section.

\section{Empirical Validation and Experiments}
\subsection{Factor Validation}
To evaluate the accuracy of our implementation, we statistically compare the computed mispricing factors implemented via Factor Engine against those generated by a reference Stata implementation. This validation involves calculating the Pearson correlation coefficient between the outputs of both systems for each factor. The reference values from Stata were generated using a different data source than the one used for Factor Engine (but for the same underlying assets and time ranges). Nonetheless, the results in Table \ref{tab:validation_new} show high Pearson correlations, confirming the correctness of the implementations.

\begin{table}[H]
\centering
\small
\caption{Validation vs. Stata Implementation}
\label{tab:validation_new}
\begin{tabular}{lc}
\toprule
\textbf{Factor} & \textbf{Pearson Correlation} \\
\midrule
CEI (1-year) & 0.9883 \\
Momentum (6mo) & 0.9787 \\
Net Operating Assets & 0.8536 \\
Accruals & 0.9334 \\
Asset Growth & 0.8077 \\
Investment to Assets & 0.8528 \\
Gross Profitability Premium & 0.9436 \\
O-Score & 0.8975 \\
\bottomrule
\end{tabular}
\end{table}
\subsection{ML-Based Trading Strategies}
Another robust approach to validating the correctness of computed mispricing factors is through the construction and evaluation of trading strategies. By using both the Factor Engine-generated factors and those from the reference Stata implementation as inputs to identical portfolio formation and backtesting procedures, one can directly compare their practical performance.

To assess the predictive power and practical utility of each factor, we first conducted a series of backtests using single-factor trading strategies. In this approach, each factor was used independently to generate trading signals and construct portfolios, allowing us to evaluate the performance of each factor in isolation. This step provides a clear view of which individual signals are most effective for forecasting returns and managing risk.

For each factor, stocks were ranked daily based on their factor scores. In the long-only version of the strategy, positions were allocated to stocks with the most favorable factor values (e.g., highest or lowest, depending on the factor’s interpretation), with equal weighting among selected stocks. In the long/short version, the strategy simultaneously took long positions in stocks with the most positive signals and short positions in those with the most negative signals, again using equal weights.

Transaction costs were incorporated by applying a fee proportional to the turnover in portfolio positions. The strategy’s performance was tracked over time by compounding daily returns, with capital set to zero if cumulative losses exceeded the initial investment (i.e., no leverage or negative capital allowed). This framework enabled a fair comparison of each factor’s standalone effectiveness, as measured by metrics such as total return, annualized return, volatility, Sharpe ratio, and maximum drawdown.

Below, we present the backtest results over a 10 year period for the top 5 factors with the highest Sharpe ratios (assuming 5\% zero-coupon rate). We also provide the S\&P500 index as a benchmark to compare with.

\begin{figure}[H]
    \centering
    \includegraphics[scale=0.27]{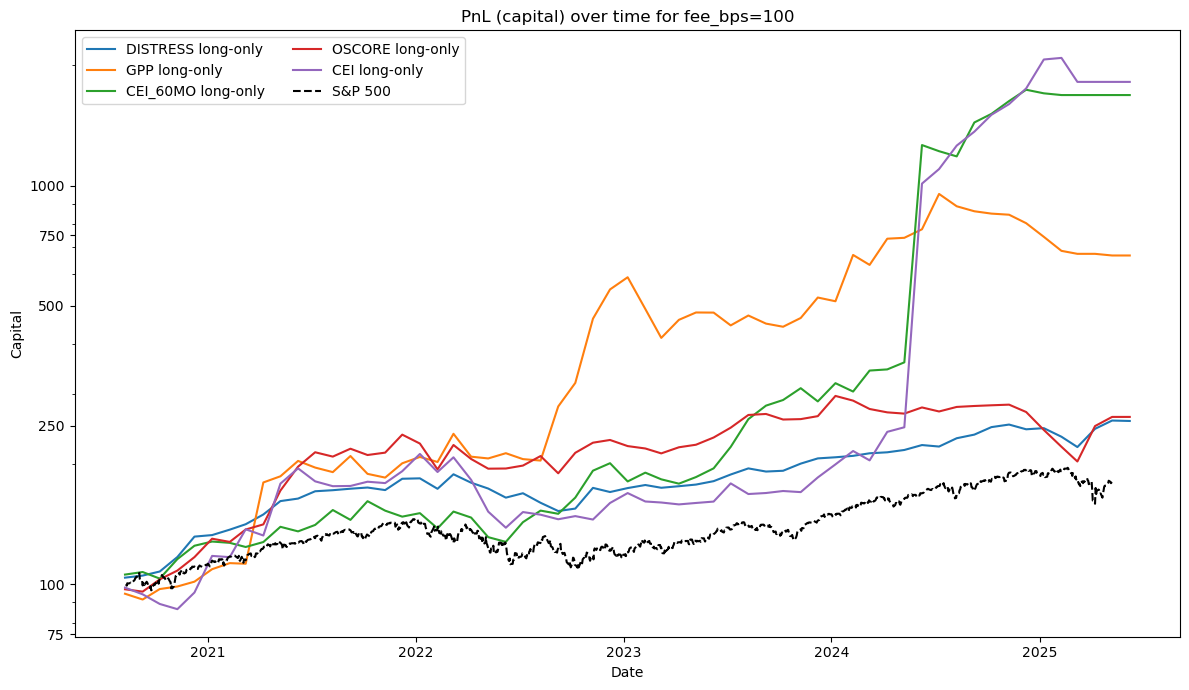}
    \caption{Single-Factor Backtest Results}
    \label{fig:100_fee_bps_top5_sharpe}
\end{figure}

Comparing this with the plot below, which is the same backtest using the factors generated by the Stata implementation, we observe that the Factor Engine factors yield similar performance metrics, confirming their reliability and correctness. In both cases the Composite Equity Index calculated over 5 years outperforms the other

\begin{figure}[H]
    \centering
    \includegraphics[scale=0.27]{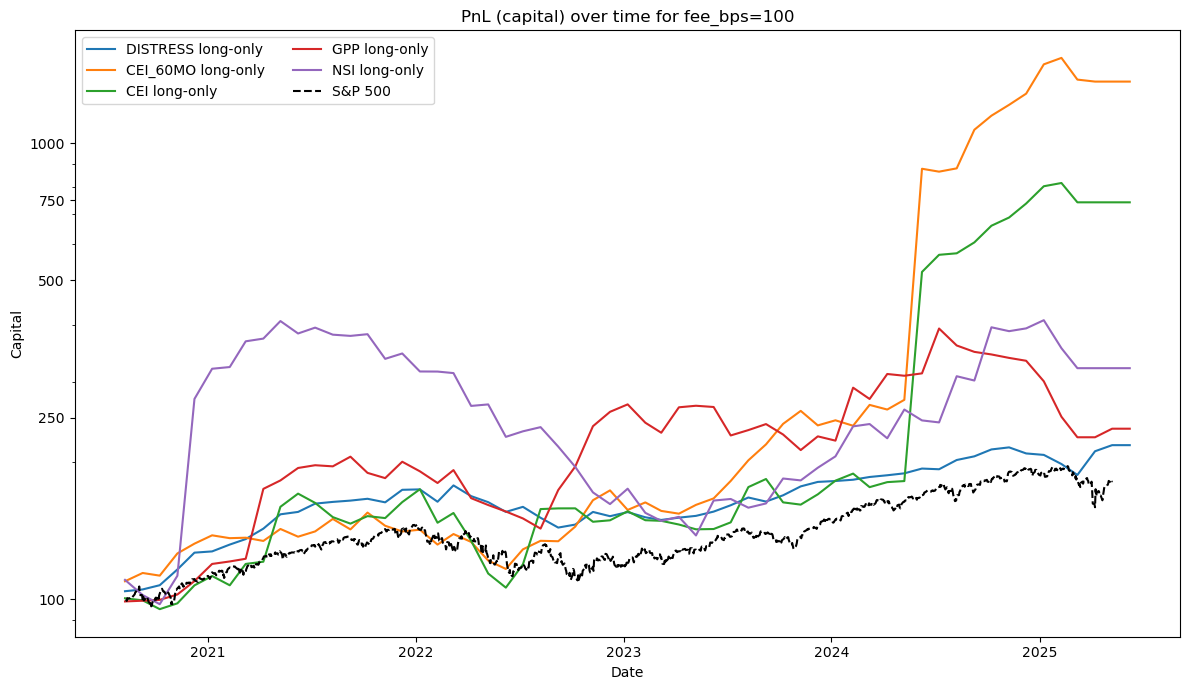}
    \caption{Single-Factor Backtest Results (Stata)}
    \label{fig:100_fee_bps_top5_sharpe_stata}
\end{figure}

Subsequently, we explored an ensemble approach by training machine learning models (such as XGBoost, LightGBM, and CatBoost) on the full set of computed factors. These models learn to combine the information from multiple factors, potentially capturing interactions and nonlinearities that single-factor strategies cannot. The ensemble-based strategies were then backtested to compare their performance against the single-factor benchmarks. For brevity, the results of the ensemble benchmarks are provided in the accompanying repository.

\section{Limitations and Future Work}
While Factor Engine is robust and flexible, its current scope is intentionally limited. The library focuses on a single suite of mispricing factors and is primarily designed for batch processing of historical data. This approach was chosen to demonstrate the core concepts and provide a practical foundation for further experimentation, rather than to serve as an exhaustive solution for all quantitative finance needs.

Future work could include expanding the factor library to cover additional families, such as the Fama-French factors, quality, and low-volatility signals, as well as improving integration with real-time data APIs for live trading. For users with very large datasets or advanced modeling requirements, optional GPU acceleration and distributed computation could be explored.

An especially promising direction is the development of AI-assisted tools for automatically generating factor code from natural language descriptions. With Factor Engine's modular architecture and decorator-based API, such a tool could translate a user's written statement of a financial factor directly into executable Python code. The clear separation between factor logic and engine mechanics means that the generated code would be concise, readable, and immediately compatible with the rest of the library. In contrast, without a framework like Factor Engine, automatic code generation would be much more challenging, as the tool would need to handle all the technical details of data alignment, lag management, and integration from scratch. By providing a standardized, extensible interface, Factor Engine makes it feasible to build intelligent systems that bridge the gap between financial domain expertise and practical implementation.

\section{Conclusion}
Factor Engine is a significant contribution to the open-source quantitative finance ecosystem. It provides a performant, transparent, and extensible solution for the task of financial factor computation. By simplifying this process, it enables practitioners to more effectively develop and test factor-based investment strategies.

\end{document}